\documentclass[a4paper]{article}
 \usepackage{graphicx}     

\usepackage{amssymb}

\begin{document}
\tolerance=10000

\def\pni{\par \noindent}
\def\vsh{\smallskip}
\def\s{\smallskip}
\def\vs{\medskip}
\def\vvs{\bigskip}
\def\vvvs{\bigskip\medskip} 
\def\vsp{\par}
\def\vsn{\vsh\pni}
\def\cen{\centerline}
\def\ra{\item{a)\ }} \def\rb{\item{b)\ }}   \def\rc{\item{c)\ }}
\def\eg{{\it e.g.}\ } 
\hyphenpenalty=2000

\font\title=cmbx12 scaled\magstep2
\font\bfs=cmbx12 scaled\magstep1
\begin{center}
 FRACALMO PRE-PRINT  \  {\tt www.fracalmo.org}
\\
Journal of Computational and Applied Mathematics, 
\\ 
Vol 207, No 2, pp. 245-257 (2007).
\vs

\hrule
\vvs

{\bfs{The role of the Fox-Wright  functions}}
\vs

{\bfs{in fractional sub-diffusion of distributed order}}
\footnote{
This paper is based on an invited talk given
by Francesco Mainardi
at the international conference {\it Special Functions: Asymptotic Analysis and Computation}, 
which took place in 
Santander (Spain) on 4-6 July, 2005. The conference was 
organized in honor of Nico M. Temme, who celebrated his 65-th birthday in 2005.
Selected papers presented at the conference are published in this  special issue 
of JCAM with the organizers as Guest Editors:
 Amparo Gil, Javier Segura (Universidad de Cantabria, Santander)
 and Jos\'e Luis L\'opez (Universidad Pública de Navarra, Pamplona).}
\vvs

Francesco MAINARDI$^{(1)}$ and Gianni PAGNINI$^{(2)}$
\vs

$^{(1)}$ Department of Physics, University of Bologna,
 and INFN, \\
   Via Irnerio 46, I-40126 Bologna, Italy\\
   {\tt francesco.mainardi@unibo.it}
\vs  

$^{(2)}$
National Agency for  New Technologies,
 Energy and the Environment,\\ 
 ENEA, Centre  "E. Clementel", 
 \\ Via Martiri di Monte Sole 4,
 I-40129 Bologna, Italy\\
 {\tt gianni.pagnini@bologna.enea.it}
\end{center}



\begin{abstract}
The fundamental solution of the fractional diffusion equation
of distributed order in time
(usually adopted for modelling  sub-diffusion processes)
 is obtained based on its
Mellin-Barnes integral representation.
Such  solution is proved  to be related via
a Laplace-type integral to the Fox-Wright
functions. 
A series expansion is also provided
in order to point out  the distribution of
 time-scales related to the distribution
of the fractional orders.
The results of the time fractional diffusion equation
of a single order are also recalled and  then
re-obtained from the general theory.
\end{abstract}
\vs

 \vs
 


\newcommand{\be}{\begin{equation}}
\newcommand{\ee}{\end{equation}}


\def\sg{\hbox{sign}\,}
\def\sgn{\hbox{sign}\,}
\def\sign{\hbox{sign}\,}
\def\e{{\rm e}}
\def\exp{{\rm exp}}
\def\ds{\displaystyle}
\def\dis{\displaystyle}
 \def\q{\quad}    \def\qq{\qquad} 
\def\lan{\langle}\def\ran{\rangle}
\def\lt{\left} \def\rt{\right}  
\def\l{\left} \def\r{\right}  
\def\lra{\Longleftrightarrow}
\def\d{\partial}
\def\dr{\partial r}  \def\dt{\partial t}
\def\dx{\partial x}   \def\dy{\partial y}  \def\dz{\partial z}
 \def\rec#1{\frac{1}{#1}}
\def\zr{z^{-1}}



\def\hatt{\widehat}
\def\epsilons{{\widetilde \epsilon(s)}}
\def\sigmas{{\widetilde \sigma (s)}}
\def\fs{{\widetilde f(s)}}
\def\Js{{\widetilde J(s)}}
\def\Gs{{\widetilde G(s)}}
\def\Fs{{\wiidetilde F(s)}}
 \def\Ls{{\widetilde L(s)}}
\def\L{{\mathcal L}} 
\def\F{{\mathcal F}} 
\def\M{{\mathcal M}}  
\def\P{{\mathcal{P}}} 
\def\H{{\mathcal{H}}} 

\def\NN{{\bf N}}
\def\RR{{\bf R}}
\def\CC{{\bf C}}
\def\ZZ{{\bf Z}} 


\def\I{{\mathcal I}}  
\def\D{{\mathcal D}}  



\def\arg{\hbox{arg}\,}
\def\erf{\hbox{erf}}  \def\erfc{\hbox{erfc}}


\def\uks{{\widehat{\widetilde {u}}} (\kappa,s)}

\def\psikappa{\psi_\alpha^\theta(\kappa)}



\section{Introduction}

The Wright function   is defined by the  series  representation,
 valid in the whole complex plane,
  $$ W _{\lambda ,\mu }(z ) :=
   \sum_{k=0}^{\infty}\frac{z^k}{k!\, \Gamma(\lambda  k + \mu)}\,,
 \q \lambda  >-1\,, \,\q \mu \in \CC\,, \q z \in \CC\,.
\eqno(1.1)
$$
It is an entire function of order $1/(1+\lambda)$,
that has been known also as {\it generalised Bessel function}\footnote{
When $\lambda =1$ the Wright function can be expressed in terms of the
Bessel function of order $\nu =\mu -1$. In fact we have
$ J_{\mu -1}(z) = \left({z}/{2}\right)^{\mu -1}\,
     W _{1 ,\mu }\left(- {z^2}/{4}\right)\,.$}.
\newpage

Originally, E.M. Wright
introduced and investigated this function
with the restriction $\lambda \ge 0$ in a series of notes starting from
1933 in the framework of the asymptotic theory of partitions
\cite{Wright 33,Wright 35a,Wright 35b}.
Only later, in 1940,
he considered the case $-1<\lambda <0\,$
 \cite{Wright 40}.
We note that in the handbook of the Bateman Project
\cite{Erdelyi HTF} (see Vol. 3,  Ch. 18),
 presumably for a misprint,  $\lambda $ is restricted to be non negative
in spite of the fact that the 1940 Wright's paper is cited.

For the cases   $\lambda > 0$ and $-1 <\lambda <0$
we agree to distinguish the  corresponding   functions
by  calling them
Wright functions  of the first and second type, respectively.
As a matter of fact the two types of functions exhibit
a quite different asymptotic behavior as it was shown
more recently in two relevant
papers by Wong  and Zaho \cite{Wong-Zaho PRS99-1,Wong-Zaho PRS99-2}.
The case $\lambda =0$ is trivial   since
it turns out from (1.1)
$W_{0,\mu}(z) = \exp(z)/\Gamma(\mu)$.

Following a former  idea of Wright himself \cite{Wright 35b},
the  Wright functions can be generalized
 as follows
$$ \, _p\Psi_q (z)
                   :=\sum_{k=0}^{\infty}
\frac{\prod_{i=1}^p \Gamma(a_i + A_i k)}
      {\prod_{j=1}^q \Gamma(b_i + B_i k)}\, \frac{z^k}{k!}
 \,, \eqno(1.2)$$
where
$ z\in \CC$,
$\{a_i,b_j\} \in \CC$,
$\{A_i,B_j\} \in \RR$
with $A_i,B_j \ne 0$ 
 and $i=1,2, \dots, p$, $j=1,2, \dots,q$.
An empty product,  when it occurs, 
is taken to be 1.
The following  alternative  notations are commonly used
$$
_p\Psi_q
\left[
 {(a_i,A_i)_{1,p} \atop   (b_j,B_j)_{1,q}}
; z \right]
= \, _p\Psi_q
\left[
{(a_1,A_1), \cdots, (a_p, A_p) \atop  (b_1,B_1), \cdots, (b_q, B_q)}
; z \right]  \,.
\eqno(1.3)$$
Then,  the standard Wright function (1.1),
being obtained from (1.2) when  $p=0$ and $q=1$
with $B_1=\lambda >-1 \,, \; b_1=\mu$,
reads
$$   W_{\lambda ,\mu }(z) \equiv  \, _0\Psi_1
    \left[
{-- \atop  (\mu  ,\lambda)}
; z \right] \,. \eqno(1.4)$$
All the above  functions  are known to belong to the more  general class of
the Fox $H$ functions introduced in 1961 by C.Fox \cite{Fox H61}.
For more information
the interested reader is referred to  the specialized literature
including the books
 \cite{Kilbas-Saigo H-BOOK04,Kiryakova BOOK94,Mathai-Saxena H-BOOK78,
Prudnikov BOOK90,Srivastava H-BOOK82},
and the relevant  articles
\cite{Kilbas FCAA05,Kilbas-Saigo H98,Kilbas-Saigo-Trujillo FCAA02,
Srivastava-Saxena-Ram AMC05}.
In particular,  we recommend the article by Kilbas et al.
\cite{Kilbas-Saigo-Trujillo FCAA02}
where the authors have established  the conditions
for the existence of  $\,_p\Psi_q (z)$, see  there \S 2, 
and provided its representations in terms of  Mellin-Barnes integrals, \S 3, 
and    Fox $H$ functions, \S 4. 

For the sake of reader's convenience  we devote
 the  Appendix for  a short outline of the $H$ functions
 in order to  understand  the Fox representation of the standard and
 generalized Wright functions of the first and second
 type  that we shall introduce in the following.
 More appropriately, following
\cite{Srivastava-Saxena-Ram AMC05,Craven-Csordas JMAA06},
we can refer to the generalized Wright functions simply  to as the
Fox-Wright functions.


The Fox notation  for the
 standard  Wright functions depends on  their  type
 and reads
  $$ W _{\lambda ,\mu }(z ) :=
   \sum_{k=0}^{\infty}\frac{z^k}{k!\, \Gamma(\lambda  k + \mu)}
 = \cases{
{\ds H^{1,0}_{0,2}
\left[-z
\left|
{-\hfill ; \hfill - \atop
(0,1);(1-\mu,\lambda)}
\right.
\right]} \,, \;
\lambda > 0 \,; \cr\cr
{\ds H^{1,0}_{1,1}
\left[-z
\left|
{-\hfill ;(\mu,-\lambda) \atop (0,1);-}
\right.
\right]} \,,  \;  -1 < \lambda < 0 \,.
}
\eqno(1.5) $$
Putting  $(b_1,B_1) =(\mu ,\lambda)$, we have 
for the  generalized Wright function:
$$\,_p\Psi_q  (z)
=\sum_{k=0}^{\infty}
\frac{\prod_{j=1}^p \Gamma(a_j + A_j k)}
{\Gamma(\mu + \lambda k) \prod_{j=2}^q \Gamma(b_j + B_j k)}
\frac{z^k}{k!} \,,
\eqno(1.6)
$$
$$
_p\Psi_q(z)= \cases{
{\ds H^{1,p}_{p,q+1}
\left[-z
\left|
\hfill{(1-a_j,A_j)_{1,p}\hfill ;\hfill - \hfill \atop
(0,1);(1-\mu,\lambda), (1-b_j,B_j)_{2,q}}
\right.
\right] \,,} \; \lambda > 0\,; \cr\cr
{\ds H^{1,p}_{p+1,q}
\left[-z
\left|
{(1-a_j,A_j)_{1,p};(\mu,-\lambda) \atop
(0,1);(1-b_j,B_j)_{2,q}}
\right.
\right]} \,,\;
-1 < \lambda < 0\,.
}
\eqno (1.7)$$
In this paper we shall show  the key-role
of the standard and generalized Wright functions of the second
type 
for finding the fundamental solutions of
diffusion-like equations containing fractional derivatives
in time of order $\beta <1$.
In the physical literature, such equations  are in general referred
to as {\it fractional sub-diffusion equations},
since they are used as model equations for the kinetic
description of  anomalous diffusion processes of  slow type,
characterized by a sub-linear grow  of the variance
(the mean squared displacement) with time.
For an easy introduction to anomalous diffusion and
fractional kinetics
 see the popular articles
\cite{Klafter-Sokolov PHYSICSWORLD05,Sokolov-et-al PHYSICSTODAY02}.

In addition to the simplest case of
a single time-fractional derivative,  more generally
we can have
  a weighted  (discrete or continuous) spectrum
of time-fractional derivatives of distributed order (less than 1):
then we speak about {\it fractional sub-diffusion of distributed order}.
We note that only from a few years the  fractional diffusion equations
of distributed order have been
investigated,  over all to describe processes of
{\it super-slow diffusion}. These processes are characterized
 by a variance  growing as a power
of the logarithm of time rather than as a linear
combination of  powers with  exponent less than 1.

We shall devote Section 2 to the simplest
case of  the time-fractional diffusion equation
of a single order: here
we  discuss how to obtain   the fundamental
solution that will be  expressed in terms of a (standard) Wright function
of second type.
In Section  3  we shall consider  the
equations of distributed order.
Starting from a  generic 
distribution of fractional derivatives,
we provide some  representations of  the fundamental solution
involving  Fox-Wright functions of the second type.
In Section 4,
as a check of consistency, we derive the fundamental
solution for the single order as a particular case of the
general representations.
Finally, the main conclusions are drawn in Section 5.

\section{The time-fractional diffusion equation \\ of single order}

It is well known that the
 fundamental solution
(or {\it Green function}) of the standard diffusion equation
$$ \frac{\d} {\dt} u(x,t) =  \frac{\d^2}{\dx^2}\,
  u(x,t)\,, \q  x\in \RR,\q t\ge 0,
\eqno(2.1)$$
 i.e. the solution
subjected to the initial condition
$ u(x, 0) = \delta (x)$ (the generalized Dirac function
\footnote{{\sc Remark}:
Through this paper  we are working {\it formally}
in that we  assume  a suitable space of generalized functions
where it is possible to deal at the same time
with delta  functions,  integral  transforms of
Fourier, Laplace, Mellin type, and fractional integrals and derivatives.}),
is
the Gaussian  {\it probability density  function} ($pdf$)
$$ u (x,t)
 = \frac{1}{2\sqrt{\pi }}\,t^{-1/2}\, \e^{-\ds x^2/(4t)}\,,
\eqno(2.2)$$
that evolves in  time with second moment\footnote{
The {centred second moment}
 provides the  {variance}  usually denoted by $\sigma^2(t)$.
It is  a  measure for the spatial spread of  $u(x,t)$ with time
of a random walking particle starting at the origin $x=0$,
pertinent to the solution of the
diffusion equation (2.1) with initial condition $u(x,0) =\delta (x)$.
The asymptotic behaviour of the variance as $t\to \infty$
is relevant to distinguish {\it normal diffusion} ($\sigma^2(t)/t \to c > 0$)
from anomalous processes of {\it sub-diffusion} ($\sigma^2(t)/t \to 0$)
and of {\it super-diffusion} ($\sigma^2(t)/t \to +\infty $).}
growing linearly with time,
$$  \mu _{2}(t) := \int_{-\infty}^{+\infty} \!\!\! x^{2}\,
  u(x,t)  \,dx = 2t\,. \eqno(2.3)$$
We note the {\it scaling property}  of the Green function,
expressed by the equation
$$ u(x,t) = t^{-1/2} \,  U(x/t^{1/2})\,, \q \hbox{with}\q
   U(x) := u(x,1) \,. \eqno(2.4)$$
The function $U(x)$ depending on the single variable $x$
turns out to be  an even function of $x$,
that is $U(x) = U(|x|)$,
and is called the {\it reduced Green function}.
The positive variable  $X :=|x|/t^{1/2}$
is known as the similarity variable.

By replacing in the  standard diffusion equation (2.1)
the first-order time derivative
by  an {\it integro-differential} operator
interpreted as a  time fractional derivative of  order $\beta \in (0,1]$,
we  obtain a   generalized diffusion equation,
the parabolic character of which is preserved.
We call it
 the {\it time-fractional diffusion equation of order $\beta $}
and, consistently  with (2.1), we write it  as
$$  
   \frac{\d^\beta }{\d t^\beta}   \, u(x,t) \,       =
\,
    \frac{\d^2  }{\d x^2 }   \,u(x,t) \,,
\q  x\in \RR,\q t\ge 0, \q 0<\beta \le 1\,,
\eqno(2.5) $$
with $ u(x,0) = \delta  (x)$.

\noindent
In (2.5)
 ${\d^\beta }/{\d t^\beta}$     denotes
the fractional  derivative (of Caputo type) of order $\beta$,
whose definition
is more easily understood if given
in terms of  Laplace transform. 
Let $f(t)$ be a sufficiently well-behaved (generalized)  function
on $t \ge 0$ with Laplace transform
$ \hbox{L} \left\{ f(t);s\right \}= \widetilde f(s)
 = \int_0^{\infty} \e^{\ds \, -st}\, f(t)\, dt$.
We have
$$  \hbox{L} \left\{ \frac{d^\beta }{ d t^\beta} f(t);s\right \} =
    s^\beta \, \widetilde f(s) - s^{\beta-1}\, f(0^+)
\q \hbox{with}\q  0<\beta \le 1,
 \eqno(2.6) $$
if we define:
$$ \frac{d^\beta}{d t^\beta} \,f(t) :=
\cases{
      {\ds \frac{1}{ \Gamma(1-\beta )}\,\int_0^t
 \frac{df(\tau )}{ d\tau }\, \frac{d\tau}{ (t-\tau )^{\beta}}}
  & for $ \; 0<\beta <1\,, $\cr\cr
 {\ds \frac{d}{dt} f(t)}
 & for $ \; \beta =1\,. $\cr
}
 \eqno(2.7) $$
For $0<\beta <1$ we can also write the fractional derivative (2.7)
 in each of the following two forms,
$$ \frac{d^\beta}{ d t^\beta} f(t)=\frac{1 }{ \Gamma(1-\beta)}\,
   \frac{d}{d t} \left[
   \int_0^t
     \frac{f(\tau )- f(0^+)}  {(t-\tau )^{\beta}}\, d \tau \right ]\,,
\eqno(2.8) $$
   $$ \frac{d^\beta}{ d t^\beta} f(t)=
  \frac{1}{\Gamma(1-\beta )}\,\frac{d}{d t} \left[
    \int_0^t  \frac{f(\tau )}{(t-\tau )^{\beta}} \,d \tau \right]
  -  f(0^+)\, \frac{t^{-\beta }}{ \Gamma(1-\beta)} \,.
\eqno(2.9)$$
We refer to the fractional derivative defined by (2.7)  as
the {\it Caputo} fractional derivative, since it was formerly
applied by Caputo in the late sixties
for modelling  dissipation effects in {\it Linear Viscoelasticity},
see  \eg
\cite{Caputo 67,Caputo 69,CaputoMaina 71,Mainardi CISM97}.
The reader should observe that Caputo's definition
differs from the usual one named after
Riemann  and Liouville, which is given by the first term in
the RHS of (2.7),
see \eg \cite{Butzer-Westphal 00,SKM 93}. For more details
we refer e.g. to \cite{GorMai CISM97,Kilbas-et-al BOOK06,Podlubny 99}.


Returning to Eq. (2.5), its fundamental solution can be obtained
by applying in sequence
the Fourier and Laplace transforms to the equation itself
\footnote{
The time-fractional diffusion equation was investigated by using
Mellin transforms
by Schneider \& Wyss \cite{SchneiderWyss 89}
in their pioneering 1989 paper where they adopted the
equivalent integral form
$$    u(x,t) =  u(x,0)    +
  {\ds{1\over \Gamma(\beta )}}\,
  {\ds \int_0^t}
 \l[{\ds \,{\d^2\over \dx^2}}\,u(x,\tau )\r] \,
  {\ds {d\tau \over (t-\tau)^{1-\beta}}}  \,.   $$
The time-fractional diffusion equation 
with the Caputo derivative has been adopted and investigated by several
authors.
From the former contributors let us quote  Mainardi, see \eg
\cite{Mainardi WASCOM93,Mainardi AML96,Mainardi CHAOS96,Mainardi CISM97}
(see also \cite{GoLuMa 99,GoLuMa 00,GorMaiSri PLOVDIV98,MainardiPagnini AMC03}
 and references therein),
who has expressed the fundamental solution in terms of a
special function (of Wright type) of which he has studied the analytical
properties and provided plots also for $1<\beta <2$.}.

Let $f(x)$ be a sufficiently well-behaved (generalized)  function
on $x \in \RR $ with Fourier transform
$ \hbox{F} \left\{ f(x);\kappa \right\} = \hat f(\kappa)
  = \int_{-\infty}^{+\infty} \e^{\,\ds i\kappa x}\,f(x)\, dx\,,
  \;\kappa \in \RR\,. $
We have
$$  \hbox{F} \left\{ \frac{d^2}{ d x^2} f(x);\kappa \right \} =
    - \kappa ^2 \, \widehat f(\kappa)
 \eqno(2.10) $$
and for the Dirac generalized function $\delta (x)$ we have
$\widehat \delta (\kappa ) \equiv 1\,. $
Then,
in the Fourier-Laplace domain our  Cauchy problem
(2.3)
appears, after applying the formulas (2.6), (2.10),
 in the form
 $  s^\beta  \, \widehat{\widetilde{u}}(\kappa ,s) - s^{\beta -1}
    = -\kappa^2\,
   \widehat{\widetilde{u}}(\kappa ,s) \,,
$
from which we obtain
$$  \widehat{\widetilde{u}}(\kappa ,s)
   =  \frac{ s^{\beta -1}}{s^\beta + \kappa ^2 }\,, \q 0<\beta \le 1\,, \q
   \q \Re (s) > 0\,,\q \kappa \in \RR\,. \eqno(2.11)$$
 To determine the  Green function
(that is expected to be symmetric in $x$)   
in the space-time domain we can follow two
alternative  strategies related to the  different
order in carrying out the
inversion of the Fourier-Laplace transforms
in (2.11).
\\
(S1) : invert  the Fourier transform
getting $\widetilde{u} (x,s)$
   and   then invert this Laplace transform;
\\
(S2) : invert  the Laplace transform
getting
$\widehat{u} (\kappa ,t)$
and then invert this Fourier transform.

{\it Strategy (S1):}
Recalling
the Fourier transform pair,
$$  {a  \over  b + \kappa ^2}
  \,\stackrel{\F} {\leftrightarrow}\,
{a \over 2 b^{1/2} }\,\e^{\ds - |x| b^{1/2}}\,,
 \q b >0 \,,
 \eqno (2.12)$$
and setting $a = s^{\beta -1}\,,\, b= s^\beta$
we get
$$ \widetilde{u} (x,s) =
 {s^{\beta /2- 1} \over 2 }\,\e^{\ds - |x| s^{\beta /2}}\,,
\q 0<\beta \le 1\,.
 \eqno(2.13) $$
The strategy (S1) has been followed by Mainardi
\cite{Mainardi WASCOM93,Mainardi AML96,Mainardi CHAOS96,Mainardi CISM97}
to obtain the  Green function  in the form
$$ u (x,t) = t^{-\beta /2}\,
       U \l(|x|/t^{\beta /2}\r)\,,
\q -\infty < x <+\infty\,,  \q t\ge 0\,,
\eqno(2.14)
$$
where the variable
 $ X:= |x|/t^{\beta /2}$ acts as {\it similarity variable}
and the function $U(x) := u(x,1)$ denotes
the {\it reduced Green function} that is  expressed in terms
of a Wright function  of the second type. Indeed we have
$$ U(x)= \frac{1}{2}\, M_{\frac{\beta}{2}} (|x|)
   = \frac{1}{2} W_{-\frac{\beta}{2} , 1-\frac{\beta}{2} }(-|x|)\,
\,, \eqno (2.15)$$
where  the $M$ function 
of order $\beta/2$  
has been introduced and investigated
in \cite{Mainardi WASCOM93,Mainardi AML96,Mainardi CHAOS96,Mainardi CISM97},
see also \cite{Podlubny 99}.
More generally,   in the complex plain
the function $M_{\frac{\beta}{2}}  (z)$  is    well
 defined for any  $\beta  \in (0,2)$ and  $\forall z \in \CC$
by  a power series as
$$
\begin{array}{ll}
 M_{\frac{\beta}{2}}  (z)
& = 
   {\ds \sum_{k=0}^{\infty}\,
  \frac{(-z)^k}{ k!\,\Gamma[-\beta  k/2 + (1-\beta /2)]}}  \\
& = {\ds \frac{1}{\pi}\, \sum_{k=0}^{\infty}\,\frac{(-z)^k }{ k!}\,
  \Gamma[(\beta  (k+1)/2]  \,\sin [(\pi \beta (k+1)/2]}
\,.
\end{array}
\eqno(2.16)$$
By comparing the power series in (1.1) and (2.16)
we recognize that
the $M_{\frac{\beta}{2}}  $ function  is indeed a special case of the
 Wright function  of the second type with $\lambda =-\beta /2$
and $\mu =1-\beta/2  $,
so that it is an entire function of order $1/(1-\beta /2)$.
Noteworthy special cases of this functions are
$$ M_{\frac{1}{2}}(z)
 = \rec{\sqrt{\pi}}\, \exp \l(-{\,z^2/ 4}\r)\,,\q
  M_{\frac{1}{3}}(z) =
  3^{2/3} \, {\rm Ai} \l( {z/ 3^{1/3}}\r) \,,    \eqno(2.17)
$$
 where {Ai} denotes the {\it Airy function},
see e.g. \cite{AS 65,Temme BOOK96}.

{\it Strategy (S2):}
Recalling
the Laplace transform pair,
see e.g.    \cite{Erdelyi HTF,GorMai CISM97,Podlubny 99},
$$
  {s^{\beta-1} \over  s^\beta +c}
  \,\stackrel{\L} {\leftrightarrow}\,
    E_{\beta} (-ct^\beta)\,,
 \q c>0\,,   
 \eqno (2.18)$$
and setting $ c = \kappa ^2$  we get
$$ \widehat{u}(\kappa ,t) =  E_{\beta} (-\kappa ^2 t^\beta)\,,
\q 0<\beta \le 1\,,    \eqno(2.19) $$
where $E_{\beta}$ denotes the  Mittag-Leffler function\footnote{
Let us recall
that the Mittag-Leffler function $E_{\beta}(z) $ ($\beta>0$)
is an entire  transcendental function of order $1/\beta $, defined
in the complex plane by the power series
$$ E_{\beta} (z) :=
    \sum_{k=0}^{\infty}\,
   {z^{k}\over\Gamma(\beta\,k+1)}\,, \q \beta  >0\,,
 \q z \in\CC\,.  $$
Originally  
Mittag-Leffler  introduced and investigated
(in five notes from 1903 to 1905)   this function
as an instructive example of entire function that generalises
the exponential (recovered for $\beta =1$).
For more details we refer e.g. to
\cite{Djrbashian 66,Erdelyi HTF,GoLoLu 02,MaiGor JCAM00,Podlubny 99}.
Here we like to recall that, for $0<\beta < 1$
and negative argument,
$E_\beta $ preserves the   {\it complete  monotonicity}
of the exponential: indeed  it
 is represented in terms of a real Laplace transform
of a positive function,
$$ E_\beta (-t^\beta ) =
   {\ds{\sin \,(\beta \pi)\over \pi}\,
   \int_0^\infty \!  \e^{\,\ds -\sigma t}\,
   { \sigma ^{\beta  -1}\, \over
    \sigma ^{2\beta } + 2\, \sigma ^{\beta }\,\cos(\beta\pi)+1}\,d\sigma}\,,
  \q t \ge 0\,,\q 0<\beta <1\,, $$
but  decreases at infinity as a power law with exponent $-\beta $:
$  E_\beta (-t^\beta ) \sim {t^{-\beta }}/{\Gamma(-\beta) }$.
In particular,   if $\beta =1/2$
we have, for $t\ge 0$ and $t\to \infty$,
$$ E_{1/2} (-\sqrt{t}) =
    \e^{\ds \, t}\, \hbox{erfc} (\sqrt{t})
\sim 1/{(\sqrt{\pi \,t})}\,,$$
where $ \, \hbox{erfc}\,$ denotes the {\it complementary error}
function, see e.g. \cite{AS 65,Temme BOOK96}.}.

The strategy (S2) has been followed
 by Gorenflo, Iskenderov \& Luchko \cite{GoIsLu 00} and
by Mainardi, Luchko \& Pagnini
\cite{Mainardi LUMAPA01}
to obtain the  Green functions of the more general
space-time fractional diffusion equations
in terms of  Mellin-Barnes integrals.
For the time fractional diffusion equation  the reduced Green function
(2.15) now appears in the form: 
$$ U(x)
 = \rec{\pi}\int_{0}^{\infty}
\!\! \cos\,(\kappa  x)  \,
  E_{\beta} \l(-\kappa^2\r)\,
d\kappa =
  \frac{1}{2x}\, \frac{1}{2\pi i}\,
   \int_{\gamma-i\infty}^{\gamma+i\infty}
  \!\!\frac{\Gamma(1-s)}{\Gamma (1- \beta s/2)}
 \, x^{\,\ds s}\,  ds
 \eqno (2.20) $$
with $0 <\gamma< 1$ and $x>0$.
We point out that from now on we restrict our attention to $x>0$
in view of the symmetry of the solution. 

In conclusion,
we may represent the solution $U(x)$ given in (2.15) and in (2.20)
using the general formalism of the Fox-Wright functions,
that is in terms of
a generalized Wright function $\,_p\Psi_q$ \cite{GoIsLu 00},
or  in terms of a Fox  $H$ function \cite{Mainardi JCAM05},
  as follows
$$ U(x) =
   \frac{1}{2}\, _0\Psi_1
    \left[
{--   \atop (1 - \frac{\beta}{2},-\frac{\beta}{2}) }
; -x \right] =
 \frac{1}{2}\,
H^{1,0}_{1,1}
\left[ x
\, \left\vert
{
{-- \;; (1 - \frac{\beta}{2},\frac{\beta}{2})\,}
\atop
{(0,1)\, ;\hfill --\hfill}
}
\right . \right]
 \,. \eqno(2.21)$$
As proven in \cite{Mainardi LUMAPA01}  we recall that
$u(x,t)$   can
interpreted
as a symmetric spatial $pdf$ evolving in time,
with a stretched exponential decay. More precisely, we have
$$  U(x) = \rec{2} \, M_{\frac{\beta}{2}} (|x|)  \sim
A \, x^a \, \e^{\ds\, -b x^c} \,, \q
x \to +\infty\,, \eqno(2.22) $$
with
$$ A=\l \{ 2\pi (2-\beta)\, 2^{\beta/(2-\beta)}
\beta^{(2-2 \beta)/(2-\beta)} \r\}^{-1/2} \,,\eqno(2.23)$$
$$ a={2\beta-2 \over 2(2-\beta)}\,, \q
b=(2-\beta)\,
 2^{-2/(2-\beta)}\beta^{\beta/(2-\beta)}\,, \q
c={2 \over 2-\beta} \,.\eqno(2.24)$$
Furthermore the moments (of even order)  of
$u(x,t)$ are
$$ \mu _{2n}(t) := \int_{-\infty}^{+\infty} \!\!\! x^{2n}\,
  u(x,t)  \,dx =
 {\Gamma(2n+1)\over \Gamma(\beta   n+1)}  t^{\beta n}\,,
\q n=0,1,2,\dots, \; t\ge 0\,.
\eqno(2.25)$$
Of  particular interest is the evolution of the  second moment:
from (2.25) we have
$$\mu _2 (t)  = 2\, \frac{t^\beta }{\Gamma(\beta +1)}\,,
\q 0<\beta \le 1\,, \eqno(2.26)$$
so that that for $\beta <1$ we note a  sub-linear growing in time,
consistently  with   an anomalous  process   of  {\it  slow diffusion}
(alternatively called {\it sub-diffusion}),
in contrast with the law (2.3) of normal diffusion.
Such result can  also  be obtained in a simpler way
from the Fourier transform (2.19) noting that
$$   \mu _2 (t)=
   -{\ds  \frac{\d^2}{\d \kappa^2}\,
 \widehat{u}(\kappa=0,t)}\,. \eqno(2.27)$$

\section{The time-fractional diffusion equation \\ of distributed order}


The fractional diffusion equation (2.5) can be generalized by using
the notion of  fractional derivative of distributed order
in time\footnote{
We find a former idea of fractional derivative of distributed order
in time in the 1969 book by Caputo \cite{Caputo 69},
that was later developed by Caputo himself,
see  \cite{Caputo FERRARA95,Caputo FCAA01}
and by Bagley \& Torvik, see \cite{BagleyTorvik 00}.}.
We now consider the so-called
{\it  time-fractional diffusion equation of distributed order}
$$  
  \int _0^1 b(\beta )\,\l[\frac{\d^\beta }{\d t^\beta}\, u(x,t)\r] \, d\beta
\,    =
\frac{\d^2  }{\d x^2 }   \,u(x,t)\,,
\q  b(\beta)\ge 0, \;  \int _0^1 b(\beta )\,d\beta =  1\,,
\eqno(3.1) $$
with $  x\in \RR, \; t\ge 0$.
 Clearly, some special conditions of regularity and behaviour near
      the boundaries will be required for the weight function
       $b(\beta)$.

Time-fractional diffusion equations of  distributed order
have  recently been discussed
in \cite{ChechkinGorenfloSokolov PRE02,ChechkinGorenfloSokolovGonchar FCAA03,%
ChechkinKlafterSokolov EUROPHYSICS03,SokolovChechkinKlafter 04}
and in \cite{Naber 04}.
As usual we consider the initial condition $u(x,0) =\delta (x)$
 in order  to keep the probability meaning.
Indeed, already in  the  paper
  \cite{ChechkinGorenfloSokolov PRE02}
it was shown that the  Green function
is non-negative and normalized, so
allowing interpretation as a density of the
probability at time $t$ of a diffusing particle to
be in the point $x$.
The main interest of those authors 
was devoted to  the second moment of the Green function
(the displacement variance or mean-square displacement)
in order to show the sub-diffusive character
of the related stochastic process by analysing
some interesting  cases of the weight function $b(\beta )$.

In this paper we are interested to a more general approach
involving   a generic 
distribution  $b(\beta )$
 in order to provide a general representation  of  the corresponding
fundamental solution.
By applying in sequence the Fourier and Laplace transforms
to Eq. (3,1) in analogy with the single-order case, see Eqs. (2.5) and (2.11),
we obtain,
$$  \widehat{\widetilde{u}}(\kappa ,s)
   =  \frac{ B(s)/s}{B(s) + \kappa ^2}\,,
   \q \Re (s) > 0\,,\q \kappa \in \RR\,, \eqno(3.2)$$
where
$$ B(s) =  \int _0^1 b(\beta )\, s^\beta \,d\beta  \,. \eqno(3.3)$$
 Before of trying to get the solution in the space-time domain,
 it is worth to outline the expression of its  second moment
as it can be derived from Eq. (3.2) using (2.27).
 We have
$$ \widehat{\widetilde{u}}(\kappa,s) =
\frac{1}{s} \, \l (1 - \frac{\kappa^2}{B(s)} + \dots \r),
\; \hbox{so} \;
\widetilde{\mu _2}(s) =
- \frac{\d^2}{\d \kappa^2} \widehat{\widetilde{u}}(\kappa=0,s)=
  \frac{2}{s\, B(s)}.\eqno(3.4)$$
Then, from (3.4)  we are allowed to  derive the asymptotic behaviours
of $\mu _2(t)$ for $t\to 0^+$ and $t\to +\infty$
from the asymptotic behaviours of $B(s)$ for $s \to \infty$
and $s \to 0$, respectively, in virtue of the Tauberian theorems.

The expected  sub-linear growth  with time is shown in the
following special cases
of $b(\beta )$ treated in
 \cite{ChechkinGorenfloSokolov PRE02,ChechkinGorenfloSokolovGonchar FCAA03}.

The first case is {\it slow diffusion} (power-law growth)
where
$$ b(\beta ) = b_1 \delta (\beta -\beta _1)+ b_2 \delta (\beta -\beta _2),
\; 0<\beta_1 <\beta _2 \le 1, \; b_1>0, \; b_2 >0,\;
b_1 +b_2 =1.$$
In fact
$$   \widetilde{\mu _2}(s) = 
\frac{2}{b_1\,s^{\beta_1+1} + b_2 \,s^{\beta_2+1}},
\q \hbox{so} \q
\mu _2(t)  \sim \cases{
 {\ds \frac{2}{b_2 \Gamma(\beta _2+1)}t^{\beta _2}}, &  $t \to 0,$ \cr\cr
{\ds \frac{2}{b_1 \Gamma(\beta _1+1)}t^{\beta _1}}, &  $t \to \infty.$}
\eqno(3.5)$$
In \cite{ChechkinGorenfloSokolov PRE02},
see  Eq. (16),
the authors were able to provide the analytical expression of
$\mu _2(t)$ in terms of a 2-parameter Mittag-Leffler function.

The second case is  {\it super-slow  diffusion} (logarithmic growth)
where
$$ b(\beta ) = 1, \q  0\le \beta \le 1\,.$$
In fact
$$   \widetilde{\mu _2}(s) =   
2\,\frac{\log s}{s(s-1)},
\q \hbox{so} \q
\mu _2(t)  \sim \cases{
 2t\, \log (1/t), &  $t \to 0,$ \cr\cr
2\, \log (t),  &  $t \to \infty.$}
\eqno(3.6)$$
In \cite{ChechkinGorenfloSokolov PRE02},
see  Eqs. (23)-(26),
the authors were able to provide the analytical expression of
$\mu _2(t)$ in terms of an exponential integral  function.

Let us now return to Eq. (3.2).
Inverting the Laplace transform, in virtue of
the Titchmarsh theorem we obtain
$$   \widehat {u}(\kappa, t) = -\rec{\pi }\,
\int_0^\infty \e^{-rt} \,  \hbox{Im}
\lt\{\widehat{\widetilde {u}} \lt(r \e^{i\pi }\rt)
\rt\} \, dr\,, \eqno(3.7)$$
that requires the expression of
$ - \hbox{Im} \lt\{
B(s)/[s(B (s)+ \kappa ^2]\rt\}$
along the ray $s=r\,\e^{i\pi }$ with $r>0$
(the branch cut of the functions   $s^\beta $ and $s^{\beta -1}$).
By writing
$$ B \lt(r\, \e^{\,\ds i \pi}\rt) =
\rho \,\cos (\pi \gamma)
+ i \rho \sin (\pi \gamma) \,,
\; \;
\cases{
{\ds  \rho =\rho (r) =\left\vert B\lt(r \,\e^{i\pi }\rt) \right\vert}\,, \cr
{\ds \gamma = \gamma (r) =
\rec{\pi}\,\arg  \left[B\lt (r \,\e^{i\pi }\rt)\right]}\,,
} \eqno(3.8)$$
 after simple calculations  we get
$$  \widehat {u}(\kappa, t) =
\int_0^{\infty}\,
\frac{\e^{\, \ds -rt}}{r} \,K  (\kappa,r) \,dr \,, \eqno(3.9)$$
where
$$ K(\kappa,r )
=    \frac{1}{\pi }\,
\frac{ \kappa ^2\rho \,\sin (\pi \gamma)}{\kappa^4 +
2 \kappa ^2 \, \rho\,\cos (\pi \gamma) +\rho^2}
\,. \eqno(3.10)$$

Then, since $u(x,t)$ is symmetric in $x$,
 the inversion formula for the Fourier transform yields
for $x,t \ge 0$,
$$
u(x,t)
=\frac{1}{\pi}\int_{0}^{+\infty} \cos(\kappa x) \,
\left\{\int_0^{\infty}
\frac{\e^{\, \ds -rt}}{r} \, \,K(r,\kappa )\, dr\right\} \, d\kappa\,.
\eqno(3.11)$$
To carry out the above Fourier integral we use the method of
the Mellin transform.
Let
$$
   {\M} \, \{ f(\xi  ); s\} = f^*(s)=
   \int_0^{+\infty} f(\xi)\,
 \xi ^{\, \ds s-1}\,  d\xi ,  \q  \gamma_1< \Re (s) <\gamma_2,
                          \eqno(3.12)
$$
be the Mellin transform of a sufficiently well-behaved function
$f(\xi)\,,$ and let
$$
{\M}^{-1}\, \{  f^*(s ); \xi  \} =f(\xi )=
{1\over 2\pi i}\int_{\gamma -i
\infty}^{\gamma +i\infty} f^*(s)\, \xi ^{\, \ds -s} \,ds\,, \eqno(3.13)
$$
 be the inverse Mellin transform,
where $\, \xi  >0\,,$ $\, \gamma = \Re (s) \,,$
$\, \gamma_1< \gamma <\gamma_2\,.$
Denoting by   $\stackrel{{\M}} {\leftrightarrow}$
the juxtaposition of a function $f(\xi )$ with its
Mellin transform $f^*(s)\,,$ the Mellin convolution
implies
$$ h(\xi  ) =  f(\xi ) \otimes g(\xi ) :=
  \int_0^\infty
 \rec{\eta}\, f(\eta )\,g(\xi /\eta  )\,{d\eta  }
\,\stackrel{{\M}}{\leftrightarrow}\,
 h^*(s) = f^*(s)\,g^*(s) \,.\eqno(3.14) $$
Then, following  \cite{Mainardi LUMAPA01} (pp. 160-161),
 we recognize that the  Fourier integral in (3.11)
 can
be interpreted as a  Mellin convolution in $\kappa $ ,
that is $u(x,t) =  f(\kappa, t) \otimes g(\kappa ,x)$,
if we set
(see  (3.14) with
  $\xi = 1/x$, $\eta =\kappa $)
 $$f(\kappa, t) :=
 \int_0^{\infty}\, \frac{\e^{\, \ds -rt}}{r} \, K(\kappa ,r) dr
       \,\stackrel{{\M}}{\leftrightarrow} \, f^*(s ,t)\,,
\eqno(3.15)$$
$$ g(\kappa,x ):= \rec{\pi\, x\, \kappa } \cos \l( \rec{\kappa}\r)
       \,\stackrel{{\M}}{\leftrightarrow} \,
{\Gamma(1-s)\over \pi \, x}
  \sin \l( {\pi s\over 2}\r)
 :=g^*(s,x)
 \,, \eqno(3.16) $$
with $0< \Re (s) <1$.
The next step thus consists in  computing the Mellin transform
$f^*(s,t)$ of the function $f(\kappa,t)$ and then  inverting the product
$ f^*(s,t) \, g^*(s,x)$ using  (3.16) in the inversion Mellin formula,
namely
$$
\begin{array}{ll}
u(x,t) 
& =
{\ds \frac{1}{\pi x}
\frac{1}{2 \pi i} \int_{\sigma-i\infty}^{\sigma+i\infty}
f^*(s,t) \, \Gamma(1-s)\sin(\pi s/2) \, x^s \,ds}  \\
& = {\ds \frac{1}{x}
\frac{1}{2 \pi i} \int_{\sigma-i\infty}^{\sigma+i\infty}
f^*(s,t) \,
\frac{\Gamma(1-s)}{\Gamma(s/2)\Gamma(1-s/2)} \,x^s \,ds}
\,.
\end{array}
    \eqno(3.17)
$$
The required Mellin transform $f^*(s;t)$  is
$$
f^*(s,t)=\int_0^{\infty} \frac{\e^{-rt}}{r}\,
\left\{   \frac{1}{\pi}\,
\int_0^{\infty}
\frac{\kappa ^2\, \rho \sin(\pi\gamma)}
{\kappa ^4+2 \rho \cos(\pi \gamma ) \kappa ^2 + \rho^2} \,
\kappa ^{\, \ds s-1}\, d\kappa
\right\} dr
\,.
\eqno(3.18)$$
The term in braces     can be computed
by making the variable change  $\kappa ^2 \to \rho \mu$
and reads
$$
\frac{\rho^{s/2+1}}{2 \rho} \,\frac{1}{\pi}
\int_0^{\infty}
\frac{\sin (\pi \gamma)}{\mu^2 + 2 \mu \cos(\pi \gamma) +1}
\, \mu^{(s/2+1)-1}\, d\mu  \qq \qq \qq \qq \qq \qq
\eqno(3.19)$$
$$ \qq \qq \qq =   -
\frac{\rho^{s/2}}{2}
\left\{
\frac{\Gamma(s/2+1)\,\Gamma[1-(s/2+1)]}
{\Gamma(\gamma s/2)\,\Gamma(1- \gamma s/2)}
\right\}
\,,
$$
where we have used a formula of the Handbook by   Marichev,
see \cite{Marichev 83} p. 156, Eq. 15 (1),
under the condition
$0< \Re (s/2+1) < 2$, $|\gamma| < 1$.
As a consequence of (3.18)-(3.19)
 we finally get
$$ f^*(s,t)=-\int_0^{\infty} \frac{\e^{\,\ds -rt}}{r}\,
   \frac{\rho^{\, \ds s/2}}{2} \,
\left\{
\frac{\Gamma(s/2+1)\,\Gamma[1-(s/2+1)]}
{\Gamma(\gamma s/2)\,\Gamma(1- \gamma s/2)}
\right\} \, dr
\,,\eqno(3.20)
$$
Now, using Eqs (3.17) and (3.20)  we can finally write the solution   as
$$
u(x,t)=\frac{1}{2\pi x}
\int_0^{\infty} \frac{e^{-rt}}{r}
F(\rho^{1/2}x)\, dr\,,   \eqno(3.21)
$$
where
$F(\rho^{1/2}x)$ is expressed in terms of Mellin-Barnes integrals:
$$
\begin{array}{ll}
F(\rho^{1/2}x)
= &
{\ds \frac{1}{2 \pi i} \int_{\sigma-i\infty}^{\sigma+i\infty}
\frac{\pi \Gamma(1-s)}
{\Gamma(\gamma s/2)\Gamma(1- \gamma s/2)}
(\rho^{1/2}x)^s \, ds}  = \\
& {\ds \frac{1}{2 \pi i} \int_{\sigma-i\infty}^{\sigma+i\infty}
\Gamma(1-s) \sin(\pi \gamma s/2)
(\rho^{1/2}x)^s\, ds} \,,
\end{array}
        \eqno(3.22)
$$
 and $\rho =\rho (r)$, $\gamma =\gamma (r)$, we remind it,
  are related
to the distribution $b(\beta )$
according to Eqs (3.3) and (3.8).
By solving the  Mellin-Barnes integrals by the residue theorem
we arrive at the series
representations in powers of $(\rho^{1/2}x)$,
$$
\begin{array}{ll}
F(\rho^{1/2}x)
& =
 {\ds \pi \rho^{1/2} x \,      \sum_{k=0}^{\infty}
\frac{(-\rho^{1/2}x)^k}
{k! \Gamma(\gamma k/2 +\gamma/2)
\Gamma(-\gamma k/2 +1 -\gamma/2)}}  \\
& = {\ds  \rho^{1/2} x \, \sum_{k=0}^{\infty}
\frac{(-\rho^{1/2}x)^k}
{k!} \sin(\pi \gamma (k+1)/2)}
 \,.
\end{array}
\eqno(3.23)
$$
Then, in virtue of Eqs (1.2)-(1.3) we recognize
$$
F(\rho^{1/2}x)= \pi \rho^{1/2} x \,\,
_0\Psi_2
\left[
{-- \atop
(1-\gamma/2,-\gamma/2) (\gamma/2,\gamma/2)}
; -\rho^{1/2}x \right] \,,
\eqno(3.24)
$$
which implies that  $F(\rho^{1/2}x)$
is a {\it Fox-Wright function of the second type} (being $\gamma >0$).
The Fox representation of the function is
$$   F(\rho^{1/2}x)=
\pi H^{1,0}_{1,2}
\left[\rho^{1/2}x
\left|
{
{\hfill -- \hfill ;(1,\gamma/2) \atop(1,1);(1,\gamma/2)}
}
\right.
\right]
\,.\eqno(3.25)
$$
In conclusion, the fundamental solution
admits the (equivalent) representations:
$$
u(x,t)=\frac{1}{2}
\int_0^{\infty} \frac{e^{-rt}}{r}\, \rho^{1/2} \,
_0\Psi_2
\left[
{-- \atop (1-\gamma/2,-\gamma/2) (\gamma/2,\gamma/2)}
; -\rho^{1/2}x \right] \,dr\,,
\eqno(3.26)
$$
and
$$
u(x,t)=\frac{1}{2 x}
\int_0^{\infty} \frac{e^{-rt}}{r}
\, H^{1,0}_{1,2}
\left[\rho^{1/2}x
\left|
{
{\hfill -- \hfill ;(1,\gamma/2) \atop(1,1);(1,\gamma/2)}
}
\right.
\right]
dr
\,.\eqno(3.27)
$$
If we exchange the order of integration and summation,
we have an alternative and interesting series representation
of the fundamental solution:
$$
u(x,t)= \frac{1}{2 \pi}
\sum_{k=0}^{\infty}
\frac{(-x)^k}{k!} \,\varphi_k(t)\,, \eqno(3.28) $$
where
$$
\varphi_k(t)=
\int_0^{\infty} \frac{e^{-rt}}{r}
\sin [\pi \gamma (k+1)/2]\,
\rho^{(k+1)/2} \,dr \,,   \eqno(3.29)
$$
with $\rho =\rho (r)$, $\gamma =\gamma (r)$.


\section{The reduction to fractional sub-diffusion \\ of a single order $\nu $}

 In order to check the consistency of the
general analysis carried out in the previous Section
and to explore  directions for future work,
we find it instructive  to derive as particular cases
the results of Section 2
concerning the fractional sub-diffusion of a single order.
We now agree to denote  this (fixed) order by $\nu $
to be distinguishes from  $\beta $ used
 in the distributed order  case (as  a variable order).
This means to consider in Eq (3.1)
 the particular case
$$ b(\beta)=\delta(\beta-\nu)\,,
 \quad 0<\nu < 1 \,,\eqno(4.1)$$
so  that  $ B(s) =  s^\nu $
and  Eq,  (3.8) yields
 $$ \rho =\rho(r)=r^{\nu}\,,\quad \gamma=const=\nu \,.\eqno(4.2)$$
In this case the Eqs  (3.9)-(3.10) reduce to
$$  \widehat {u}(\kappa, t) =
\int_0^{\infty}\,
\frac{\e^{\, \ds -rt}}{r} \,K  (\kappa,r) \,dr \,, \q
 K(\kappa,r )
=    \frac{1}{\pi }\,
\frac{ \kappa ^2\rho \,\sin (\pi \nu )}{\kappa^4 +
2 \kappa ^2 \, \rho\,\cos (\pi \nu ) +\rho^2}
\,, \eqno(4.3)$$
and hence, in virtue of (3.11),
$$
u(x,t)=\frac{1}{\pi} \int_0^{\infty} \cos(kx)\,
\left\{
\frac{\sin(\pi \nu)}{\pi}\,
\int_0^{\infty}
\frac{  \kappa^2\,  r^{\nu-1}\,\e^{\,\ds -rt} }
{\kappa ^4 + 2 r^{\nu} \cos(\pi \nu) \kappa ^2 + r^{2 \nu}}
\, dr \right\}\, d\kappa
\,. \eqno(4.4)$$
With the change of variable $r= \sigma  \kappa ^{2/\nu}$
the term in brace  reads
$$
    \frac{\sin(\nu \pi)}{\pi}\,
\int_0^{\infty} \,e^{\,\ds - \sigma  \kappa ^{2/\nu} t}
  \,\frac{\sigma ^{\nu -1} }
{\sigma ^{2 \nu} + 2 \sigma ^{\nu} \cos(\nu \pi) + 1}\, d\sigma \,,
     \eqno(4.5)
$$
so that
$$ u(x,t)=\frac{1}{\pi} \int_0^{\infty} \cos(kx)\,
 E_{\nu}\lt(-\kappa ^2t^{\nu}\rt)\, d\kappa \,,\eqno(4.6)$$
where  $E_{\nu}$ is the Mittag-Leffler function of order $\nu $
according to its integral representation of  footnote $\null^{(6)}$.
We thus recognize that Eq. (4.6)  is consistent  with  Eq. (2.19),
and, once applied the scaling relation for the Fourier transform,
with Eqs. (2.14)-(2.15) and (2.20).

The consistency with the results expressed   in terms
of the general formalism of Fox-Wright function,
that is the comparison between Eqs. (2.21) and (3.26)-(3.27),
can be obtained  in a less direct  way because
one is required to  use the scaling relations
and the Laplace transform rules  of the Fox $H$ functions
available in the   specialized literature. We do not
report on this tedious calculations.
In a more direct and instructive way  the consistency
with the single-order case
is shown by using the
series representation of the fundamental solution (3.28)-(3.29).
In this special  case
the functions  $\varphi_k(t)$    turn out to be
$$
\begin{array}{ll}
 \varphi_k(t)  
 & =
   {\ds \sin[\pi \nu (k+1)/2]\,
\int_0^{\infty} \frac{\e^{\,\ds -rt}}{r}\,
 r^{\, \ds \nu (k+1)/2}\, dr}  \\
& = {\ds \sin[\pi \nu (k+1)/2]\, \frac{\Gamma [\nu(k+1)/2]} {t^{\nu(k+1)/2}}}
 \,.
\end{array}
\eqno(4.7)
 $$
As a consequence,
the solution reads
$$
\begin{array}{ll}
 u(x,t)
  & =
{\ds \frac{1}{2} t^{-\nu/2} \cdot
\frac{1}{\pi }\, \sum_{k=0}^{\infty}
\frac{(-x/t^{\nu/2})^k}{k!}\, \Gamma [\nu(k+1)/2]\,
\sin[\pi \nu (k+1)/2]} \\
 &  = {\ds \frac{1}{2} t^{-\nu/2} \,
M_{\frac{\nu}{2}} \lt( \frac{x}{t^{\nu/2}}\rt)}\,,
\end{array}
\eqno(4.8)
$$
in agreement with Eqs. (2.14)-(2.16).
Of course, only in this special  case it is
possible to single out  a common  time factor  ($t^{-\nu/2}$)
from all the functions $\varphi_k(t)$
and get a self-similar  solution.
In  general the set of  functions $\varphi_k(t)$
give raise to a distribution
of different time scales related
in some way to the distribution of the orders
of the fractional derivatives.


\section{Conclusions}
The diffusion-like  equations containing fractional derivatives
in time and/or in space are usually adopted
to model  phenomena of anomalous transport in physics, so
a detailed study of their solutions is required.
   Our attention in this paper has been focused on the
time fractional diffusion equations of
 distributed order less than 1, which are known to be   model equations
for sub-diffusive processes.
Specifically, we have worked out how express
their fundamental solutions in terms
of Fox-Wright functions.

At first we have recalled the main results  for the
fundamental solution of the
time fractional diffusion equation of a single order,
which are obtained  by applying  the Fourier-Laplace integral transforms.
The  required solution turns out to be self similar
(through a definite space-time scaling relationship),
and  expressed in terms of
a special function  belonging
 to the simpler class of the Wright functions.
Then we  were able to adapt  the previous techniques
for obtaining the fundamental solution
in the general  case of a distributed order.
For such solution  we have provided a representation
in terms of
a Laplace-type integral of a  Fox-Wright function, 
that  can be expanded in  a series containing
 powers of  space   and certain functions of time, responsible
of the time-scale distribution.

Among the    various questions for future research on this topic,
particularly relevant in our opinion is the possibility to use
our analytical results  for plotting the fundamental
solutions in some noteworthy cases of fractional  order distribution,
as it was for the simplest case of a single order.

\subsection*{Acknowledgements}
The authors are grateful to 
R. Gorenflo and the anonymous referees
for useful comments and suggestions.
\newpage

\section*{Appendix: The Fox H functions}

According to a standard notation  the Fox $H$ function is defined as
$$  H^{m,n}_{p,q} (z)= \rec{2\pi i}\,
\int_{\L} \H^{m,n}_{p,q} (s) \, z^s \, ds \,, \eqno(A.1)$$
where
 $\L$ is a suitable 
path in the complex plane  $\CC\,$
to be disposed later,
  $ z^s=\exp \{ s(\log |z| + i \, \arg z)\}$, 
and
$$
 \H^{m,n}_{p,q} (s)=  {A(s)\, B(s) \over C(s)\, D(s)}\,,
 \eqno(A.2)$$
$$  A(s) = \prod_{j=1}^m \Gamma(b_j-B_j s)\,, \q
 B(s)= \prod_{j=1}^n \Gamma(1-a_j +A_j s)\,,
 \eqno(A.3)$$
$$ C(s) = \prod_{j=m+1}^q \Gamma(1-b_j +B_j s)\,,\q
D(s)= \prod_{j=n+1}^p \Gamma(a_j-A_j s)\,.
 \eqno(A.4)$$
with
$ \, 0\le n\le p\,$, $\, 1\le m\le q\,$,
$\, \{a_j, b_j \}  \in \CC\,$,  $\,\{A _j, B _j \}  \in \RR^+\,.$
An empty product, when it occurs, is taken to be one so
   $$  n=0 \lra B(s) =1\,, \q
       m=q \lra C(s) =1\,, \q
       n=p \lra D(s) =1\,. $$
Due to the occurrence of the factor $z^s$ in the integrand of (A.1),
the $H$ function is, in general, multi-valued,
but it can be made one-valued on the  Riemann surface of $\log z$
by choosing  a proper branch.
We also note that
when  the $A$'s and $B $'s are equal to 1,
we obtain the Meijer's $G$-functions
$  G^{m,n}_{p,q}(z)$.

The above integral representation of the $H$ functions, by involving
products and ratios of Gamma functions,  is known to be of
{\it Mellin-Barnes integral} type
\footnote{ 
As historical note we point out that
 the names refer to the two authors, who in the first 1910's
 developed the theory of these integrals  using them
 for a complete integration of the hypergeometric differential equation.
 However,
these integrals were first used in 1888 by S. Pincherle,
see e.g.  
\cite{Mainardi-Pagnini JCAM03}.
Recent treatises on Mellin-Barnes integrals are those
by Marichev \cite{Marichev 83}
and Paris \& Kaminski \cite{ParisKaminski BOOK01}.
}.
 A compact  notation is usually adopted for $(A.1)$ :
$$ H^{m,n}_{p,q}(z) = H^{m,n}_{p,q}
\left[z   \left\vert
{\hfill (a_j ,A _j )_{j=1,n}; (a_j ,A _j )_{j=n+1,p}  \hfill \atop
 \hfill (b_j,B _j)_{j=1, m}; (b_j,B _j)_{j=m+1, q} \hfill}
 \right.
\right]\,.  \eqno(A.5)$$
Thus, the singular points of the kernel $\H$  are the poles of
the Gamma functions entering the expressions
of $A(s)$ and $B(s)$, that we assume do not coincide.
Denoting by ${\P}(A)$  and ${\P}(B)$ the sets
of these poles, we write
$ {\P}(A) \,  \cap {\P}(B) = \emptyset\,.$
The conditions for the existence of the $H$-functions
can be made by inspecting the convergence of the integral
(A.1), which can depend  on the selection of the contour $\L$
and on certain relations between the parameters
$\{a_i,A _i\}$ ($i=1, \dots, p$) and
$\{b_j, B _j\}$ ($j=1, \dots, q$).
For the analysis of the general case we refer to
the specialized treatises on $H$ functions, \eg
 \cite{Mathai-Saxena G-BOOK73,Mathai-Saxena H-BOOK78,Srivastava H-BOOK82}
and, in particular to the paper by Braaksma
\cite{Braaksma 62}, where an exhaustive discussion on
 the asymptotic expansions
and analytical continuation of these functions is found,
see also \cite{Kilbas-Saigo H98}.


\end{document}